\documentclass{optica-article}

\journal{opticajournal} 

\articletype{Research Article}

\usepackage{lineno}

\usepackage{colortbl}
\definecolor{mygray}{gray}{.9}

\begin{document}

\title{MEMO: Dataset and Methods for Robust Multimodal Retinal Image Registration with Large or Small Vessel Density Differences}

\author{Chiao-Yi Wang,\authormark{1,*} Faranguisse Kakhi Sadrieh,\authormark{1} Yi-Ting Shen, \authormark{2} Shih-En Chen, \authormark{3} Sarah Kim, \authormark{3} Victoria Chen, \authormark{3} Achyut Raghavendra, \authormark{3} Dongyi Wang, \authormark{4} Osamah Saeedi, \authormark{3,*} and Yang Tao \authormark{1,*}}

\address{\authormark{1}Department of Bioengineering, University of Maryland, College Park, MD 20742, USA\\
\authormark{2}Department of Electrical and Computer Engineering, University of Maryland, College Park, MD 20742, USA\\
\authormark{3}Department of Ophthalmology and Visual Sciences, University of Maryland School of Medicine, Baltimore, MD 21201, USA\\
\authormark{4}Department of Biological and Agricultural Engineering, University of Arkansas, Fayetteville, AR 72701, USA}

\email{\authormark{*}cyiwang@umd.edu; OSaeedi@som.umaryland.edu; ytao@umd.edu;} 


\begin{abstract*} 
The measurement of retinal blood flow (RBF) in capillaries can provide a powerful biomarker for the early diagnosis and treatment of ocular diseases. However, no single modality can determine capillary flowrates with high precision. Combining erythrocyte-mediated angiography (EMA) with optical coherence tomography angiography (OCTA) has the potential to achieve this goal, as EMA can measure the absolute RBF of retinal microvasculature and OCTA can provide the structural images of capillaries. However, multimodal retinal image registration between these two modalities remains largely unexplored. To fill this gap, we establish \textit{MEMO}, the first public multimodal EMA and OCTA retinal image dataset. A unique challenge in multimodal retinal image registration between these modalities is the relatively large difference in vessel density (VD). To address this challenge, we propose a segmentation-based deep-learning framework (\textit{VDD-Reg}), which provides robust results despite differences in vessel density. \textit{VDD-Reg} consists of a vessel segmentation module and a registration module. To train the vessel segmentation module, we further designed a two-stage semi-supervised learning framework (\textit{LVD-Seg}) combining supervised and unsupervised losses. We demonstrate that \textit{VDD-Reg} outperforms existing methods quantitatively and qualitatively for cases of both small VD differences (using the CF-FA dataset) and large VD differences (using our \textit{MEMO} dataset). Moreover, \textit{VDD-Reg} requires as few as three annotated vessel segmentation masks to maintain its accuracy, demonstrating its feasibility.
\end{abstract*}

\section{Introduction}
\label{sec:introduction}
Retinal blood flow (RBF) is a key functional biomarker, implicated in three of the four major causes of blindness worldwide, glaucoma \cite{nicolela1996ocular}, diabetic retinopathy \cite{patel1992retinal}, age-related macular degeneration \cite{ciulla1999color}, as well as in neurodegenerative diseases such as Alzheimer's dementia \cite{feke2015retinal, berisha2007retinal}. Specifically, RBF in capillaries may provide sensitive biomarkers for the early diagnosis of ocular diseases, and could aid in the development of novel therapies. Unfortunately, accurately measuring RBF in capillaries is challenging because it requires the precise measurement of both absolute erythrocyte velocities and capillary width. Moreover, it also has high requirements of sensor resolution and repeatability.

Current methods of measuring RBF are limited. For instance, laser Doppler imaging \cite{riva2010ocular} is limited by high variability of measured flowrates. Dynamic OCTA \cite{lee2017face, jia2012split} and color Doppler imaging \cite{goebel1995color} can only measure relative flowrates, leading to poor intra-platform and cross-platform measurement repeatability. Adaptive optics scanning laser ophthalmoscopy (AO-SLO) \cite{roorda2010applications, arichika2013noninvasive} and AO-OCT \cite{pircher2017review, carroll2013adaptive} have a limited field of view. Erythrocyte mediated angiography (EMA) \cite{flower2008observation}, on the other hand, is a novel technique which has the capability of determining absolute erythrocyte flowrates of arterioles and venules \textit{in vivo} with high precision and a large field of view. EMA determines the flowrates by following the motion of individual fluorescently labeled erythrocyte ghosts in the retinal capillary circulation which can be visualized \textit{in vivo} \cite{saeedi2018determination, tracey2019measurement, asanad2021erythrocyte}. Despite the aforementioned advantages, a major limitation of EMA is its inability to delineate the capillary structures through which the erythrocytes are flowing \cite{wang2019automated}.

One potential solution to address this limitation of EMA is to combine it with another modality that provides high-resolution structural imaging of retinal capillaries. Optical coherence tomography angiography (OCTA) is an ideal candidate, as it can generate high resolution imaging down to the capillary level in different layers of the retina \cite{kashani2017optical, gao2016optical, watanabe2014graphics}. Combining EMA and OCTA may enable absolute capillary RBF measurement for diagnosis and treatments of ocular diseases. A key requirement for this is accurate registration. Manual approaches to registration are time-consuming, necessitating the development of an automated approach to registration of EMA and OCTA image pairs.

Multimodal retinal image registration has been extensively studied in recent years \cite{santarossa2022medregnet, de2021deep, luo2020multimodal, arikan2019deep, lee2019deep, zhang2021two, wang2021robust, sindel2022multi}. However, current approaches have primarily utilized the public CF-FA dataset \cite{hajeb2012diabetic} (color fundus and fluorescein angiography) or private datasets with modalities other than EMA and OCTA, such as CF and fundus autofluorescence (FAF) \cite{santarossa2022medregnet, de2021deep}, and CF and infrared reflectance (IR) imaging \cite{de2021deep, zhang2021two, wang2021robust}. The lack of new and publicly available multimodal retinal image datasets not only makes it difficult for researchers to fairly and thoroughly compare existing methods, but also prevents the identification of novel methods of multimodal imaging registration.

To fill in these gaps in knowledge, we conducted experiments on non-human primates (NHPs) and created a public dataset of EMA and OCTA pairs. NHPs are used extensively in ophthalmic research and provide some of the best models for glaucoma as well as other ocular diseases \cite{burgoyne2015non}. The homology between NHPs and human eyes allows for ease in translation of experimental findings and applicability to human imaging and disease. Our dataset has the unique features of being well-controlled, one of the few datasets that includes OCTA images and the only dataset to include EMA sequences. This dataset is described as the \textbf{m}ultimodal \textbf{EM}A and \textbf{O}CTA (\textit{\textbf{MEMO}}) retinal image dataset. \textit{MEMO} contains EMA and OCTA image pairs with manually labeled matched points for studying multimodal retinal image registration. Additionally, \textit{MEMO} includes OCTA projection images \cite{rocholz2019oct} from all three retinal vascular plexi (superficial vascular plexus (SVP), intermediate capillary plexus (ICP) and deep capillary plexus (DCP)) and EMA image sequences.

\begin{figure}[htbp]
\centerline{\includegraphics[width=\columnwidth]{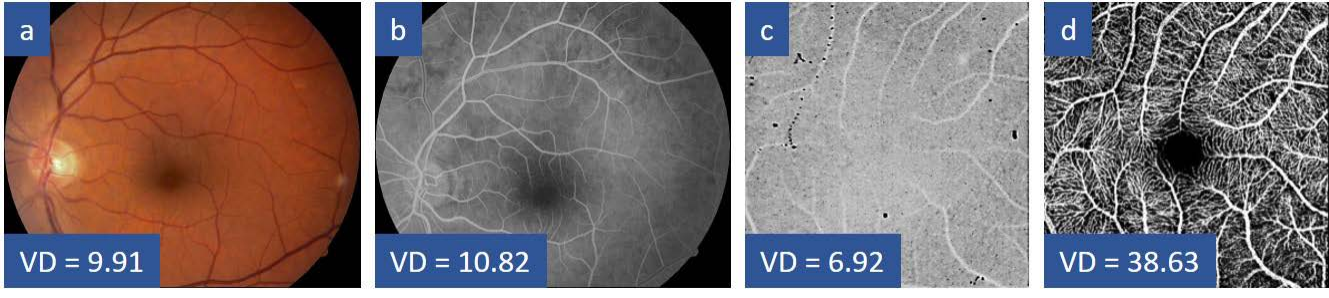}}
\caption{Sample images of (a) CF, (b) FA, (c) EMA and (d) OCTA with vessel density (VD). (a) and (b) are taken from the CF-FA dataset \cite{hajeb2012diabetic}. In this example, the vessel density of OCTA (d) is five times grater than that of EMA (c) since most capillaries cannot be visualized in EMA images.}
\label{figIntro2}
\end{figure}

Using the \textit{MEMO} dataset, we address a unique challenge in multimodal retinal image registration between EMA and OCTA images arising from the relatively large difference in vessel density between the two modalities. In this paper, the vessel density (VD) is defined as the proportion of image area occupied by vessels divided by the entire captured area. As compared to other modalities available in public datasets, such as CF-FA \cite{hajeb2012diabetic}, EMA and OCTA have a VD difference over 30\% between the two modalities (Fig.~\ref{figIntro2}). Through extensive experiments, we found that large VD differences dramatically decrease registration performance, as the majority of smaller vessels in OCTA could not be visualized in EMA due to fundamental differences in image acquisition.

To overcome the challenges posed by large VD differences, we propose \textit{\textbf{VDD-Reg}}, a segmentation-based deep-learning framework for multimodal retinal image \textbf{reg}istration that can robustly register two imaging modalities despite \textbf{v}essel \textbf{d}ensity \textbf{d}ifferences. \textit{VDD-Reg} consists of a vessel segmentation module and a registration module. Here, instead of trying to extract every vessel as accurately as possible, the goal of the segmentation module is to extract vessels that are visible in both modalities so that the registration module can detect and match feature points more accurately. To achieve this goal, we designed a novel two-stage semi-supervised learning framework, \textit{\textbf{LVD-Seg}}, which requires only a few (e.g., three) labeled vessel segmentation masks from the modality with lower vessel density (EMA in our case). Specifically, \textit{LVD-Seg} first uses a supervised loss (i.e., MSE) to stabilize the training of the vessel segmentation module, and then uses an unsupervised loss (i.e., style loss \cite{johnson2016perceptual}) with a unified style target image to guide the segmentation module to extract common vessels visible in both EMA and OCTA images, improving the registration accuracy.

The contributions of our work can be summarized as follows:
\begin{enumerate}
    \item We establish \textit{MEMO}, the first public multimodal EMA and OCTA retinal image dataset. \textit{MEMO} provides registration ground truth, all three retinal vascular plexi of OCTA projection images, and EMA image sequences containing moving erythrocytes. This also has the potential for use in any research involving OCTA registration with other modalities that use a scanning laser ophthalmoscope. \textit{MEMO} is the first retinal image dataset containing EMA images and also the first multimodal retinal image registration dataset containing modalities with a large difference in vessel density (VD). \textit{MEMO} is available at \url{https://chiaoyiwang0424.github.io/MEMO/}.
    
    \item We propose a segmentation-based deep-learning framework, \textit{VDD-Reg}, for multimodal retinal image registration that is robust with respect to vessel density differences. To train the segmentation module in \textit{VDD-Reg}, we further designed a two-stage semi-supervised learning framework, \textit{LVD-Seg}, which requires as few as three labeled vessel segmentation masks. 
\end{enumerate}

The rest of the paper is organized as follows. Section~\ref{RelatedWork} summarizes the existing public retinal image datasets with image pairs and multimodal retinal image registration methods. Section~\ref{Dataset} illustrates the details of our \textit{MEMO} dataset. In Section~\ref{Method}, we describe the proposed \textit{VDD-Reg} framework. Section~\ref{ExpSettings} illustrates our experimental settings. Section~\ref{Result} and Section~\ref{Discussion} present the results and discussion. Section \ref{DisCon} includes the conclusion of the paper.

\section{Related Works} 
\label{RelatedWork}
\subsection{Retinal Image Datasets with Image Pairs}

\begin{table*}[t]
\caption{Comparison of Public Retinal Image Datasets with Image Pairs}
\label{table_dataset}
\resizebox{\columnwidth}{!}{
\begin{tabular}{cccccccc}
\hline
\multicolumn{1}{c}{Dataset} & \multicolumn{1}{c}{Multi-modal} & \multicolumn{1}{c}{Modality} & \multicolumn{1}{c}{\begin{tabular}[c]{@{}c@{}}Pair / \\ Image Number\end{tabular}} & \multicolumn{1}{c}{\begin{tabular}[c]{@{}c@{}}Global Registration\\ Ground Truth\end{tabular}} & \multicolumn{1}{c}{Initial Purpose} \\
\hline
e-ophtha \cite{decenciere2013teleophta}    & -           & Color fundus                              & 144 pairs           & -  & Diabetic retinopathy\\
\rowcolor{mygray}
VARIA \cite{ortega2009retinal}   & -           & Fundus image                              & 154 pairs           & -   & User authentication\\
RODREP \cite{adal2015accuracy}       & -           & Color fundus                              & 1400 pairs          & - & Image registration\\
\rowcolor{mygray}
FIRE \cite{hernandez2017fire}               & -           & Color fundus                              & 134 pairs           & \checkmark & Image registration\\
FLORI21 \cite{ding2021flori21}           & -           & Ultra-widefield fluorescein   angiography & 15 pairs            & \checkmark & Image registration\\
\rowcolor{mygray}
OCTA-500 \cite{li2020image}                    & \checkmark  & 1. OCTA  2. OCT                   & 500 pairs           & - & Vessel segmentation \\
PRIME-FP20 \cite{ding2020weakly} & \checkmark & \begin{tabular}[c]{@{}c@{}}1. Ultra-widefield fundus photography \\2. Ultra-widefield fundus angiography\end{tabular} & 15 pairs & - & Vessel segmentation\\
\rowcolor{mygray}
CF-FA  \cite{hajeb2012diabetic}      & \checkmark & \begin{tabular}[c]{@{}c@{}}1. Color fundus \\2. Fluorescein angiography\end{tabular}   & 59 pairs & - & \begin{tabular}[c]{@{}c@{}}Diabetic retinopathy,\\ Image registration *\end{tabular} \\
\textbf{MEMO (Our Dataset)} & \textbf{\checkmark}  & \textbf{1. EMA  2. OCTA}                  & \textbf{30 pairs}   & \textbf{\checkmark} & \textbf{Image registration}      \\   \hline
\end{tabular}}
\begin{minipage}{\columnwidth}
\vspace{0.2cm}
\scriptsize $^{*}$ The registration ground truth is not officially provided. 
\end{minipage}
\end{table*}

There are relatively few public retinal image datasets specifically curated for image registration. In Table~\ref{table_dataset}, we summarized the public retinal image datasets with image pairs as they could be potentially repurposed for image registration with proper ground truth annotations. The datasets listed in Table~\ref{table_dataset} can be divided into monomodal and multimodal. The monomodal datasets, such as e-ophtha \cite{decenciere2013teleophta}, VARIA \cite{ortega2009retinal}, RODREP \cite{adal2015accuracy}, FIRE \cite{hernandez2017fire} and FLORI21 \cite{ding2021flori21}, contain images from only one modality, limiting their utility for multimodal retinal image registration research. On the other hand, existing multimodal retinal image datasets provide images from various modalities, such as OCT-OCTA pairs \cite{li2020image}, ultra-widefield fundus photography-angiography pairs \cite{ding2020weakly}, and CF-FA pairs \cite{hajeb2012diabetic}. Compared to the above datasets, \textit{MEMO} has three major advantages. Firstly, \textit{MEMO} is the first retinal dataset with EMA images and also the first multimodal retinal image registration dataset providing two modalities with relatively large VD differences. Secondly, \textit{MEMO} officially provides six corresponding point pairs per image pair as the global registration ground truth, which is crucial for fair comparisons of methods. Finally, \textit{MEMO} provides raw EMA sequences and OCTA projection images, which may be useful for multiple research fields such as automated erythrocyte tracking.

\subsection{Multi-Modal Retinal Image Registration}
\label{related:methods}

Multimodal retinal image registration methods can be categorized into conventional and deep learning-based methods. The conventional methods can be further divided into two types: \textit{direct} and \textit{indirect} methods. The \textit{direct} conventional methods try to detect and match features directly on the raw images by manually designing more powerful feature descriptors or more robust matching algorithms. For example, Chen et al. \cite {chen2010partial} proposed a partial intensity invariant feature descriptor (PIIFD) and designed an image registration framework called Harris-PIIFD based on the proposed descriptor. Ghassabi et al. \cite{ghassabi2013efficient} combined UR-SIFT and PIIFD for image registration with large content or scale changes. Wang et al. \cite{wang2015robust} presented an image registration framework combining SURF, PIIFD and robust point matching. Lee et al. \cite{addison2015low} introduced a low-dimensional step pattern analysis method to align retinal image pairs that were poorly aligned with baseline methods. Hossein-Nejad et al. \cite{hossein2018ransac} adopted adaptive Random Sample Consensus (A-RANSAC) for feature matching. On the other hand, the \textit{indirect} conventional methods attempt to first transfer the images from different modalities into a similar "style", such as the vessel mask or the phase image, before detecting and matching features. For instance, Hernandez et al. \cite{hernandez2015multimodal} proposed line structures segmentation with a tensor-voting approach to improve registration. Hervella et al. \cite{hervella2018multimodal} combined feature-based and intensity-based registration methods and employed a domain-adapted similarity metric to detect vessel bifurcations and crossovers. Motta et al. \cite{motta2019vessel} proposed a registration framework based on optimal transport theory for vessel extraction on retinal fundus images. Li et al. \cite{li2018multi} proposed a two-step registration method which converted raw images into phase images and adopted log-Gabor filters for global registration.

Recently, many deep learning-based multimodal retinal image registration methods have been proposed, demonstrating comparable or superior performance as compared to conventional methods. Similar to the conventional methods, deep learning-based methods can also be roughly divided into \textit{direct} and \textit{indirect} methods. The \textit{direct} deep learning-based methods usually try to directly learn a feature matching network using raw image datasets. For example, De Silva et al. \cite{de2021deep} proposed an end-to-end network following the conventional feature point-based registration steps, using a VGG-16 feature extractor \cite{simonyan2014very} and a feature matching network for predicting patch displacements. Lee et al. \cite{lee2019deep} extracted pattern patches surrounding the intersection points and used a Convolutional Neural Network (CNN) to select matched patches. The \textit{indirect} deep learning-based methods, on the other hand, try to learn a transformation network to first transform the two modalities into the same domain such as the vessel mask instead of directly performing image registration. For instance, Arikan et al. \cite{arikan2019deep} used a U-Net for vessel segmentation and a mask R-CNN for vessel junctions detection based on supervised learning before multimodal image registration. Luo et al. \cite{luo2020multimodal} proposed a two-stage affine registration framework. The first stage used two individual U-Nets to segment the optic discs in two modalities, and the second stage adopted ResNet for fine registration. Zhang et al. \cite{zhang2021two} proposed a vessel segmentation-based two-step registration method integrating global and deformable registration. Their vessel segmentation networks were trained with a deformable registration network using ground truth registration affine matrices. Wang et al. \cite{wang2021robust} proposed a content-adaptive multimodal retinal image registration method, which adopted pixel-adaptive convolution (PAC) \cite{su2019pixel} and style loss \cite{johnson2016perceptual} in their vessel segmentation network. In addition to transforming images into the vessel masks, Santarossa et al. \cite{santarossa2022medregnet} and Sindel et al. \cite{sindel2022multi} applied CycleGAN \cite{zhu2017unpaired} to transform the images from one modality to the other before extracting features.

Although many methods have been proposed for multimodal retinal image registration, none of them tackle the registration between EMA and OCTA. Compared to the modalities used in existing works, the vessel density (VD) difference between EMA and OCTA used in our \textit{MEMO} dataset is relatively large, making image registration much more challenging.

\section{The MEMO Dataset} 
\label{Dataset}
\subsection{Overview}

\begin{figure*}[htbp]
\centerline{\includegraphics[width=\columnwidth]{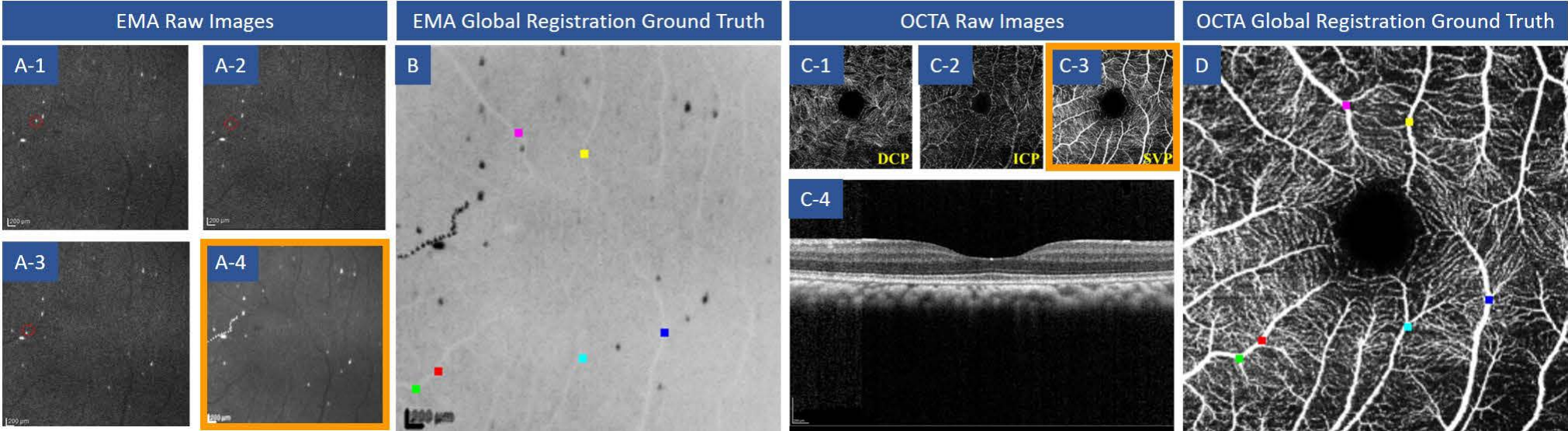}}
\caption{A typical sample EMA and OCTA pair from our \textit{MEMO} dataset. Images inside the orange boxes were used for ground truth labeling. (A-1, A-2 and A-3: frame 0, 10 and 20 in the sample EMA image sequence. A-4: the stacked images of the EMA sequence. C-1, C-2 and C-3: the sample OCTA projection images representing DCP, ICP and SVP layer. C-4: the OCTA B-scan image. B and D: the six corresponding point pairs of the sample EMA and OCTA pair.)}
\label{figDataset}
\end{figure*}

\begin{figure}[htbp]
\centerline{\includegraphics[width=0.7\columnwidth]{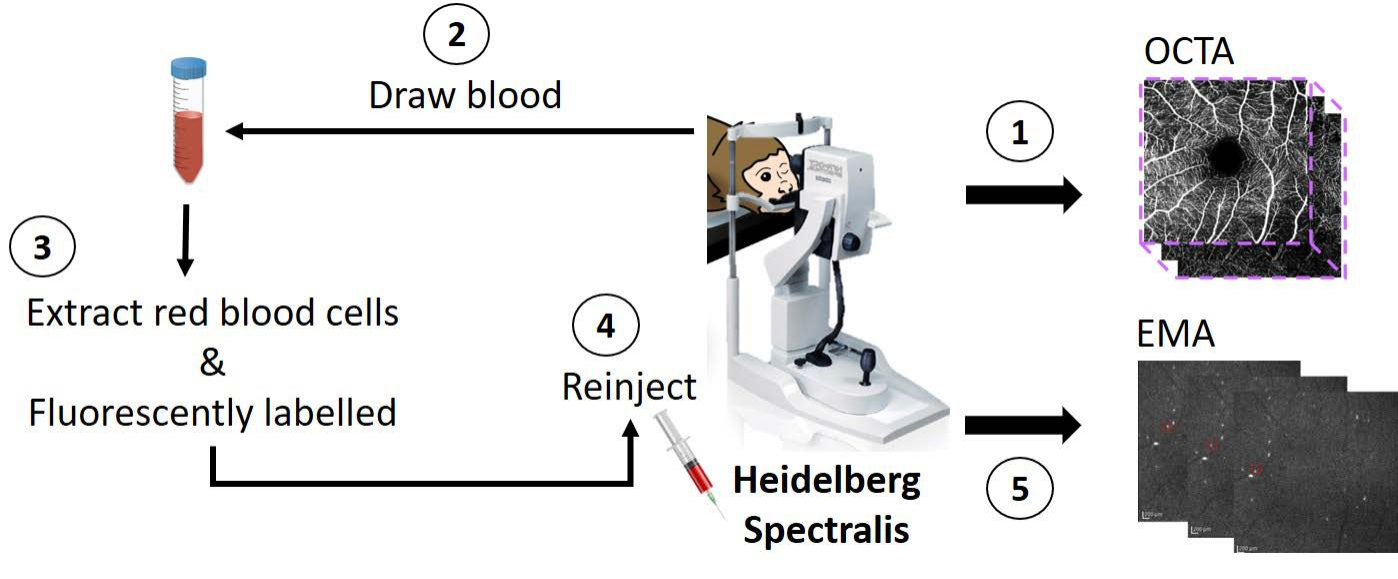}}
\caption{The procedure for image acquisition. The numbers shown in the figure indicate the order.}
\label{figExp}
\end{figure}

A sample EMA and OCTA image pair from the \textit{MEMO} dataset is shown in Fig.~\ref{figDataset}. The dataset contains 30 pairs of EMA and OCTA images. For each image pair, 6 corresponding point pairs were manually annotated. The annotated points were chosen from the visually distinctive points in EMA and OCTA images, such as vessel bifurcation points and vessel bending points. The procedure for EMA and OCTA image acquisition is shown in Fig.~\ref{figExp}. All images were acquired following a protocol approved by Institutional Animal Care and Use Committee of the University of Maryland, Baltimore. Four eyes from two healthy non-human primates (rhesus monkeys, i.e., \textit{Macaca mulatta}), aged 14 and 20, were used to acquired paired EMA and OCTA images. Each pair was collected in the same session on the same date. All image pairs were captured by a Heidelberg Spectralis platform (Heidelberg Engineering, Heidelberg Germany), which minimizes the nonlinear effect between two modalities or between different machines. Prior to the experimental session, the animal was sedated with ketamine and xylazine (5-10 and 0.2-0.4 mg/kg by intramuscular injection). The animal was intubated by trained veterinary technicians with an endotracheal tube and general anesthesia was maintained with 1.5\% to 3\% isoflurane with 100\% oxygen. The animal was paralyzed with vecuronium (40-60 ug/kg, followed by 0.35-45 ug/kg/min), preventing eye movement during image acquisition. Body temperature was maintained at physiologic levels using a thermal blanket and blood pressure was monitored using a blood pressure cuff on the arm. The animal was laid in a prone position during the imaging session. A wire lid speculum was used to keep the eyelids open during imaging and tropicamide 1\% was administered for pupillary dilation.

\subsection{EMA}

All EMA image sequences were captured by a Heidelberg Spectralis platform (Heidelberg Engineering, Heidelberg Germany). Each image sequence represents a time sequence capturing the trajectory of a single moving erythrocyte as it travels from the artery through the capillary to the vein. Approximately 17 mL of blood was drawn for processing with 5,6-carboxyfluorescein diacetate succinimidyl ester (CFSE) (Molecular Probes, USA) reconstituted in anhydrous dimethyl sulfoxide with a method similar to the human erythrocyte preparation, as previously described in our previous work \cite{tracey2019measurement}. Autologous erythrocytes were isolated from whole blood and loaded with 7.5 mM of CFSE using the osmotic shock method. Following cell preparation, up to 1.2 mL of CFSE-loaded cells were intravenously injected during image acquisition. After the cells were injected, ten-second angiograms centered on the macula were obtained with the Heidelberg Spectralis (Heidelberg Engineering GmbH, Germany) using a high speed 15-degree horizontal x 15-degree vertical field of view taken at 15 frames per second. More details of the procedure can be referred to our published protocol \cite{tracey2019measurement, pottenburgh2020use, chen2022vivo}.

All image frames from the EMA image sequences were stored in TIF format. The number of image frames for each sequence is different, since the speed of each erythrocyte is different. In addition, six image sequences had the image size of $512\times512$ pixels, while the other 24 had the image size of $384\times384$ pixels. The reason for different sizes of image is because some images were taken in high-speed mode, while others were taken in high-resolution mode which has a higher pixel density. The pixel size of every EMA image sequence was provided. The stacked image of each EMA image sequence was used for registration ground truth labeling.

\subsection{OCTA}

OCTA scans centered on the fovea were taken using the same Heidelberg Spectralis with a $10\times10$ degree protocol, consisting of 512 a-scans $\times$ 512 b-scans with 5-10 microns between b-scans and 5-7 frames averaged per b-scan location. Projection images of the superficial vascular plexus (SVP), intermediate capillary plexus (ICP), and deep capillary plexus (DCP) were generated using the segmentation algorithms and slab definitions provided by the Spectralis software (Heidelberg Eye Explorer, version 1.10.3.0, Heidelberg Engineering, Germany). The SVP slab was defined as between the internal limiting membrane to the anterior border of the inner plexiform layer, the ICP included the entire inner plexiform layer, and the DCP ranged from the posterior border of the inner plexiform layer to the anterior border of the outer plexiform layer. The projection images were processed using projection artifact removal (PAR). 

All images from the OCTA image groups were stored in TIF format. Each of the OCTA image groups contained three images from the three layers (i.e., SVP, ICP and DCP). Fifteen OCTA image groups had the image size of $512\times512$ pixels, while the other 15 had a image size of $768\times768$ pixels. The SVP image from each OCTA image group was used for registration ground truth labeling.

\subsection{Dataset Analysis}

\begin{figure}[t]
\centerline{\includegraphics[width=0.9\columnwidth]{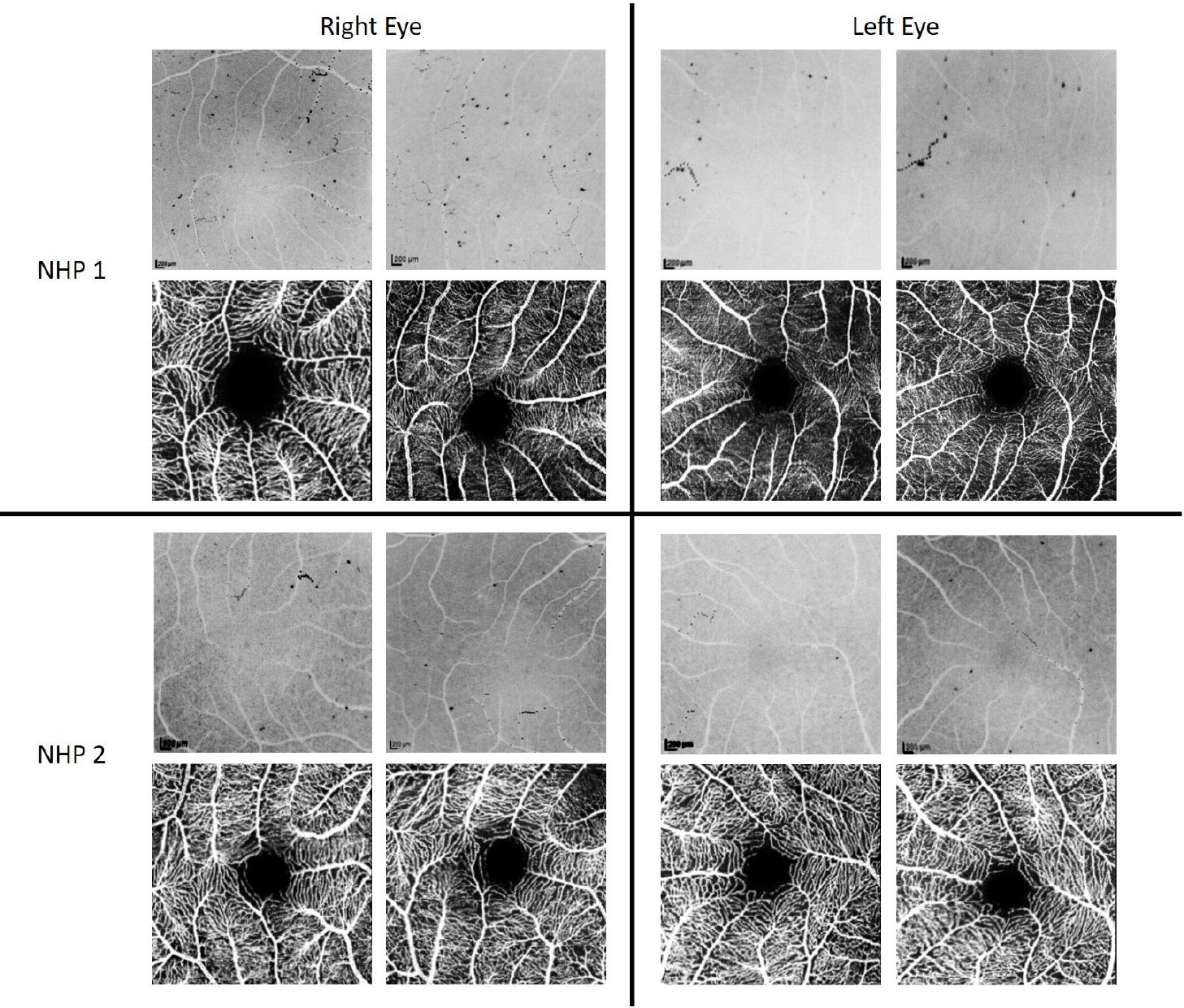}}
\caption{Image samples corresponding to each eye of each NHP. The EMA image is placed on top of the OCTA image for each image pair.}
\label{figMemoEx}
\end{figure}

\begin{figure}[t]
\centerline{\includegraphics[width=1.0\columnwidth]{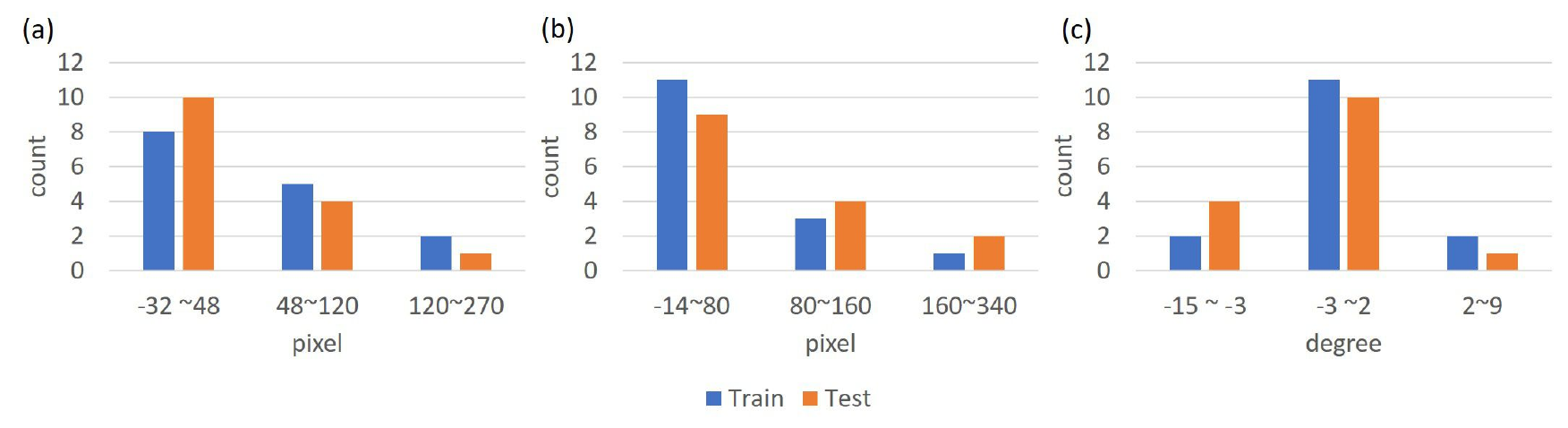}}
\caption{The Statistics of our \textit{MEMO} dataset. The number of image pairs (count) falling within different ranges of (a) translation in the x-axis (pixel), (b) translation in the y-axis (pixel), or (c) rotation (degree) are presented. The division of training and test data for the \textit{MEMO} dataset is outlined in Sec.~\ref{ExpDataset:MEMO}.}
\label{figMemoAnalysis}
\end{figure}

Despite its size, our \textit{MEMO} dataset has sufficient diversity. To showcase this, additional image pairs corresponding to each eye of each non-human primate (NHP) are shown in Fig.~\ref{figMemoEx}. The images exhibit noticeable differences from each other, even if captured from the same NHP or eye. These differences arise from varying physical conditions of the NHPs, different capture times, imaging variations, etc. Besides, we present additional statistics of our \textit{MEMO} dataset in Fig.~\ref{figMemoAnalysis}, particularly focusing on the image registration aspect. It is observed that our \textit{MEMO} dataset covers various image transformations.

\section{Proposed Method} 
\label{Method}
\begin{figure*}[t]
\centerline{\includegraphics[width=\linewidth]{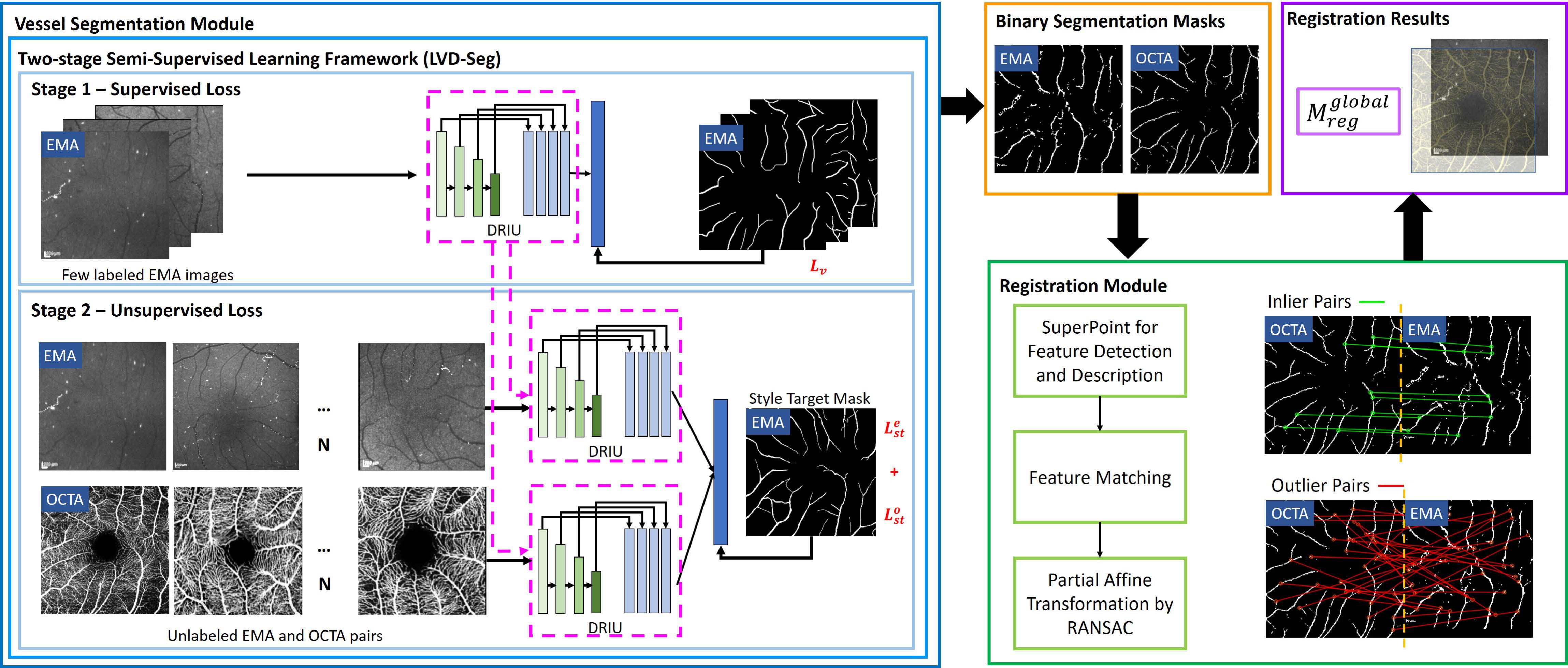}}
\caption{The proposed \textit{VDD-Reg} framework. \textit{VDD-Reg} includes a vessel segmentation module and a registration module. The vessel segmentation module is trained with the proposed two-stage semi-supervised learning framework (LVD-Seg). DRIU \cite{maninis2016deep} and SuperPoint \cite{detone2018superpoint} are adopted for our segmentation networks and registration network, respectively. $M_{reg}^{global}$ denotes the partial affine transformation matrix for global image registration.}
\label{figframework}
\end{figure*}

The overview of the proposed framework, \textit{VDD-Reg}, for multimodal retinal image registration is shown in Fig.~\ref{figframework}, which consists of a vessel segmentation module and a registration module. In \textit{VDD-Reg}, multimodal images were first transformed into binary vessel masks by the vessel segmentation module. The global registration matrix was then estimated by the registration module from the two binary vessel masks. 

\subsection{Vessel Segmentation Module}

\subsubsection{LVD-Seg Background}

As discussed in Section~\ref{related:methods}, vessel segmentation has frequently been used as the first step for multimodal retinal image registration \cite{arikan2019deep, zhang2021two, wang2021robust, hervella2018multimodal, hernandez2015multimodal, motta2019vessel}, primarily because features of vessels are considered to be more consistent across different modalities. Recently, deep learning-based vessel segmentation methods have shown superior performance. They can be categorized into two groups, supervised \cite{arikan2019deep, maninis2016deep} and unsupervised methods \cite{zhang2021two, wang2021robust}, which present different limitations. The supervised vessel segmentation methods \cite{arikan2019deep, maninis2016deep} usually require a large number of high-quality pixel-level vessel masks for training to ensure test performance, but such high-quality pixel-level vessel masks are often difficult and time-consuming to acquire. To avoid the need for pixel-level vessel masks, the unsupervised vessel segmentation methods based on style transfer have been proposed \cite{zhang2021two, wang2021robust}. However, due to the lack of direct supervision, the unsupervised vessel segmentation methods generally performs worse than the supervised ones in terms of the segmentation quality.

Unlike general vessel segmentation, which aims to accurately extract every vessel, the goal of the vessel segmentation module in our \textit{VDD-Reg} is to extract vessels that are visible in both modalities. This is particularly crucial for multimodal retinal image registration when a majority of vessels in one modality (e.g. OCTA) are not visible in the other modality (e.g. EMA) due to the fundamental differences between the two modalities. To this end, we designed a novel two-stage semi-supervised learning framework, \textit{LVD-Seg}, to train our vessel segmentation module. \textit{LVD-Seg} was designed based on two key insights. First, \textit{LVD-Seg} combined supervised and unsupervised training so that the resulted segmentation masks could be effectively used by the following registration module despite having very few pixel-level vessel masks. Second, only the pixel-level vessel masks from the modality with lower vessel density (e.g., EMA) were used as the supervisory signal for both supervised and unsupervised training, guiding the segmentation module to extract vessels that are visible in both modalities. Details of the two stages in \textit{LVD-Seg} are described as follows.

\subsubsection{LVD-Seg Stage 1 - Supervised Loss}
\label{method:supervised}

In this stage, we trained our vessel segmentation module on $n$ manually-annotated EMA vessel segmentation masks, where $n$ could be as few as three according to our experiment results (Sec.~\ref{dis:gt}). We used EMA vessel segmentation masks because the vessel density of EMA is much lower than that of OCTA and vessels that can be captured by EMA are visible in both modalities. In other words, OCTA images contain a plethora of small capillaries which do not present in the corresponding EMA images and are not helpful for image registration. Moreover, labeling the less complex EMA vessel segmentation masks is much more feasible in terms of efficiency than labeling OCTA vessel segmentation masks.

Following \cite{zhang2021two, wang2021robust}, we adopted the DRIU \cite{maninis2016deep} network for segmenting EMA images. The DRIU network used a pre-trained VGG-16 network \cite{simonyan2014very} for feature extraction and was followed by a segmentation prediction layer. The mean squared error (MSE), denoted as $L_{v}$, was adopted to train the network, which is defined as
\begin{equation}
L_{v} = \frac{1}{N} \sum_{i=1}^{N}(Pred(I(i)) - M(i))^2.
\label{eq1}
\end{equation}
where $I$ represents the input EMA image and $M$ represents the ground truth EMA mask. $Pred(I)$ represents the predicted segmentation mask of $I$. $i$ represents the $i^{th}$ pixel of the predicted segmentation mask or the ground truth mask. $N$ denotes the total number of pixels. In addition to MSE, the self-comparison loss \cite{zhang2021two, wang2021robust}, denoted as $L_{sc}$, was also adopted to make the prediction robust against input image rotation. Specifically, $L_{sc}$ is defined as 
\begin{equation}
L_{sc} = MSE(Rot_{-90}(Pred(Rot_{90}(I))), \ Pred(I)),
\label{eq2}
\end{equation}
where $Rot_{\theta}(I)$ represents $I$ rotated by $\theta$ degree. Here, $L_{sc}$ can be seen as an alternative way to perform data augmentation. Overall, the training loss for stage 1, denoted as $L_{s1}$, can be written as
\begin{equation}
L_{s1}= w_{v} * L_{v} + w_{sc} * L_{sc},
\label{eq3}
\end{equation}
where $w_{v}$ and $w_{sc}$ represent the weighting factors for the MSE and the self-comparison loss. In this paper, $w_{v}$ and $w_{sc}$ were set to 1 and 1e-3, respectively. Here, $w_{sc}$ was set to a relatively smaller value compared to $w_{v}$ to ensure that $L_{v}$, the primary supervisory signal in this stage, could effectively guide the learning process. The trained weights of the EMA vessel segmentation network in stage 1 were used to initialize both the EMA and OCTA vessel segmentation networks in stage 2.

\subsubsection{LVD-Seg Stage 2 - Unsupervised Loss}

To ensure the extraction of common vessels visible in both modalities, we further optimized the segmentation networks using style loss \cite{zhang2021two, wang2021robust, johnson2016perceptual} with a joint style target mask, a stand-alone EMA ground truth segmentation mask. Using a joint style target mask with style loss guided the EMA and OCTA segmentation networks to segment shared vessels, improving the registration accuracy. Moreover, using the same trained weights to initialize both the EMA and OCTA vessel segmentation networks further stabilized the unsupervised training in this stage.

Style loss penalized the difference between the predicted segmentation mask and the style target mask using Gram matrices \cite{gatys2015texture, johnson2016perceptual}. The Gram matrix was used to capture the style information but remove the spatial information, whose elements can be written as 
\begin{equation} 
G_{j}(I)_{c,c^{'}} = \frac{1}{C_j H_j W_j} \sum_{h=1}^{H_j} \sum_{w=1}^{W_j} \phi_{j}(I)_{h, w, c} \phi_{j}(I)_{h, w, c^{'}}.
\label{eq4}
\end{equation}
Here, $\phi_{j}(I)$ denotes the feature map with shape $C_j \times H_j \times W_j$ obtained from the $j^{th}$ layer of the pre-trained VGG-16 network \cite{simonyan2014very} by feeding the network with input image $I$, and $c, c^{'} \in [1, \ C_j]$. The Gram matrix, a $C_j \times C_j$ matrix, is utilized to capture correlations among features, thus representing the `style' of the input image $I$. Style loss ($L_{st}$) is then defined as the squared Frobenius norm of the difference between the Gram matrices of the predicted segmentation mask ($Pred(I)$) and the style target mask ($M_{t}$), which can be written as
\begin{equation}
L_{st_{j}}(Pred(I), \ M_{t}) = \parallel G_{j}(Pred(I)) - G_{j}(M_{t}) \parallel_{F}^{2},
\label{eq5}
\end{equation}
\begin{equation}
L_{st} = \sum_{j \in J} L_{st_{j}}(Pred(I), \ M_{t}).
\label{eq5_2}
\end{equation}
$I$ represents the input EMA or OCTA image. Style loss was computed at four different layers of the VGG-16 network. Overall, the training loss for stage 2, denoted as $L_{s2}$, can be written as
\begin{equation}
L_{s2}= w_{st}^{e} L_{st}^{e} + w_{st}^{o} L_{st}^{o} + w_{sc} (L_{sc}^{e} + L_{sc}^{o}).
\label{eq6}
\end{equation}
Note that the two different modalities used different weighting factors for style loss. The self-comparison loss was also adopted as an additional constraint for the predicted segmentation masks. In this paper, $w_{st}^{e}$, $w_{st}^{o}$ and $w_{sc}$ were set to 100, 1 and 1e-3, respectively. Similar to Eq.~\ref{eq3}, $w_{sc}$ was set to a relatively smaller value to ensure that the primary supervisory signals in this stage, $L_{st}^{e}$ and $L_{st}^{o}$, effectively guided the learning process. Additionally, we observed that $L_{st}^{o}$, representing the style loss for OCTA, tended to be larger than $L_{st}^{e}$ due to the use of the EMA vessel segmentation ground truth mask as the style target. To balance the impacts of $L_{st}^{o}$ and $L_{st}^{e}$, we assigned $w_{st}^{e}$ to a relatively larger value compared to $w_{st}^{o}$.

The outputs of the segmentation module were EMA and OCTA pixel-wise probability maps, which represented the probability of each pixel belonging to a vessel. The probability maps were transformed into binary segmentation masks with threshold set to 0.5.

\subsection{Registration Module}
\subsubsection{Feature Detection and Description}

\begin{figure}[htbp]
\centerline{\includegraphics[width=\linewidth]{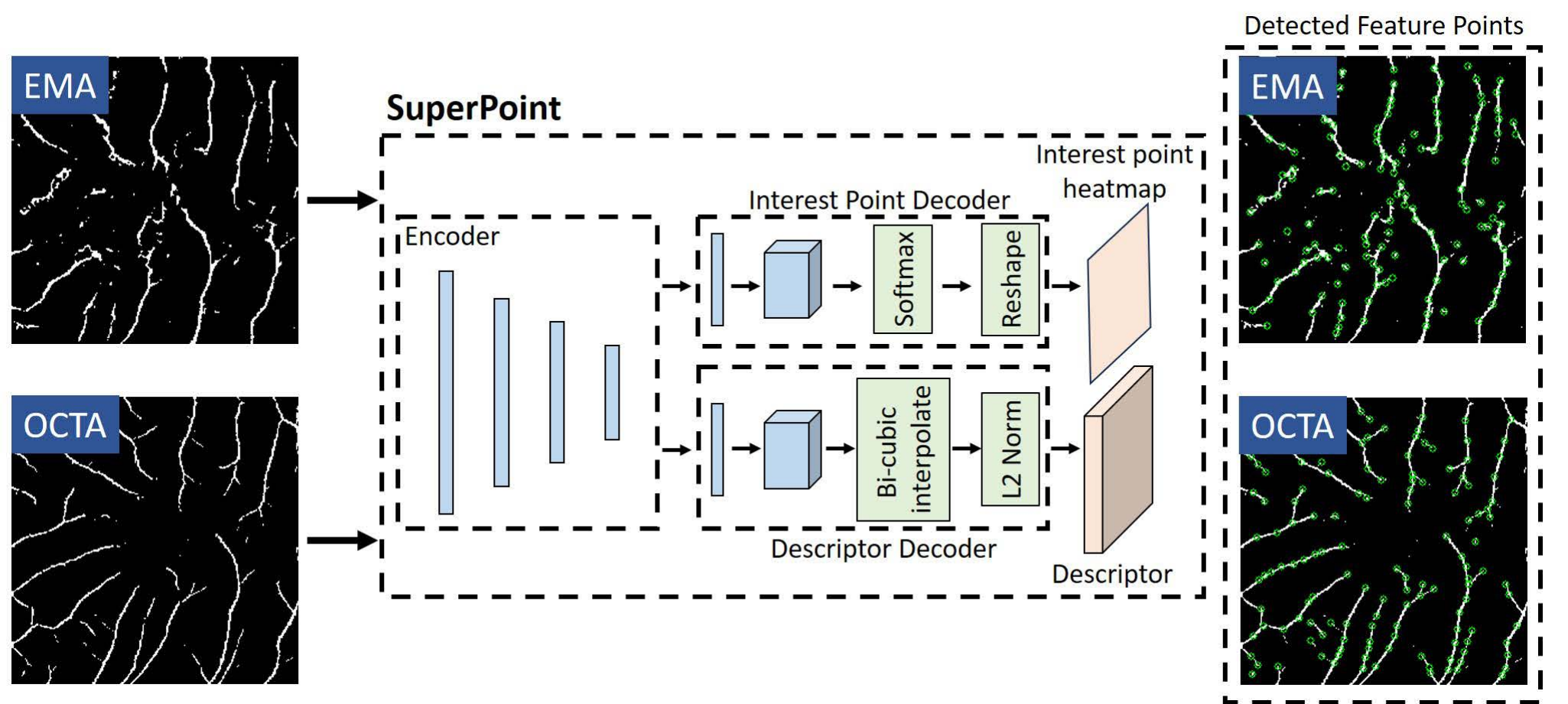}}
\caption{The network structure of SuperPoint \cite{detone2018superpoint}. The green circles indicate the feature points detected by SuperPoint.}
\label{figSuperPoint}
\end{figure}

We adopted pre-trained SuperPoint \cite{detone2018superpoint} as our feature detector and descriptor. It demonstrated superior performance than many traditional feature detectors and descriptors, and has been widely adopted in many applications which require feature matching, including multimodal image registration \cite{zhang2021two, jiang2021review, zhao2022heterogeneous}, localization \cite{sarlin2019coarse} and two-view geometry estimation \cite{zhang2019learning}. The network structure of SuperPoint is shown in Fig.~\ref{figSuperPoint}. It contains a shared encoder, an interest point decoder and a descriptor decoder. It was first trained on a synthetic dataset with labeled interest points to detect feature points. Next, Homographic Adaptation \cite{detone2018superpoint} was used to self-label a large unlabeled real world image dataset. Finally, the model was jointly-trained to extract feature points and their corresponding descriptors with self-supervision. We refer the readers to \cite{detone2018superpoint} for more details. In this paper, the non-maximum suppression distance was set to 4 and the detector confidence threshold was set to 0.015 for keypoint detection.

\subsubsection{Feature Matching and Registration}

We determined the matched feature points based on bidirectional calculation of the minimum Euclidean distance. That is, a feature point X in an OCTA image is said to match a feature point Y in the corresponding EMA image only when the Euclidean distance between each other's feature descriptors is smaller than (1) the Euclidean distance between X and any other feature point in the EMA image and (2) the Euclidean distance between Y and any other feature point in the OCTA image. The Random Sample Consensus (RANSAC) \cite{fischler1981random} method was applied to remove outliers and estimate the partial affine transformation matrix between the EMA and OCTA pair. Here, the partial affine transformation (i.e., 4 degrees of freedom) was adopted because the EMA and OCTA images in \textit{MEMO} were captured using the same device and already have the same pixel density (i.e., scale factor).

\section{Experimental Settings} 
\label{ExpSettings}
\subsection{Dataset} 
\label{ExpDataset}

We used our \textit{MEMO} dataset and the CF-FA \cite{hajeb2012diabetic} dataset to conduct the experiments. The two datasets were chosen to examine how the proposed and the competing methods performed for both scenarios of small VD differences (using the CF-FA dataset) and large VD differences (using our \textit{MEMO} dataset). 

\subsubsection{The MEMO Dataset}
\label{ExpDataset:MEMO}

The MEMO dataset contains 30 pairs of images. Fifteen pairs (with even indices) were selected as the training set, and the rest of the pairs (with odd indices) were used as the test set. For OCTA, the SVP layer projection images were used in our experiments because they contained clearer arterioles and venules, and were free of projection artifact. For EMA, the stacked image of each EMA image sequence was used for the purpose of denoising. Furthermore, we annotated the vessel segmentation mask of each EMA stacked image for \textit{VDD-Reg}. We also annotated one EMA stacked image that is not part of the \textit{MEMO} dataset as the style target. Note that even though we annotated the vessel masks for all EMA images, our \textit{VDD-Reg} actually required only three of those to maintain its performance.  

For the image pre-processing, the OCTA images were first resized to $256\times256$ pixels. Then, the EMA images were resized using the same scaling factors of the corresponding OCTA images. Next, to meet the requirement of our model, the resized EMA images were then cropped to ensure that their widths and heights were multiples of 8. Finally, the annotated EMA vessel segmentation masks and the registration ground truth were pre-processed accordingly to ensure their correct scale. To ensure the quality and consistency of annotation, all annotations were drawn by the same human annotator and checked by an experienced ophthalmologist (OJS).

\subsubsection{The CF-FA Dataset}

The CF-FA dataset contains 59 pairs of color fundus (720 $\times$ 576, RGB) and fluorescein angiography (720 $\times$ 576, grayscale) images. Twenty-nine pairs of images are from healthy subjects, while the other 30 pairs are from patients with retinopathy. We manually labeled 6 pairs of corresponding points for all image pairs as the registration ground truth. We selected 29 image pairs (with odd indices) as the training set, 29 image pairs (with even indices) as the test set, and 1 image pair (normal/1-1) as the style target. Similar to the \textit{MEMO} dataset, we manually annotated the vessel segmentation masks of the style target and three selected color fundus images from the training set for the proposed \textit{VDD-Reg}. For image pre-processing, the color fundus (CF) images were converted to grayscale, with no additional pre-processing steps applied.

\subsection{Comparison to Existing Methods}
\label{exp::baseline}

\begin{table}
\centering
\caption{Summary of Five Existing Methods}
\setlength{\tabcolsep}{3pt}
\resizebox{0.8\columnwidth}{!}{
\begin{tabular}{c|c|c|c|c}
\hline
Methods & Year    & Deep learning-based & Indirect & Transfer Method     \\
\hline
SURF-PIIFD-RPM \cite{wang2015robust}   & 2015 & -                         & -          & -                   \\
LoFTR \cite{sun2021loftr} & 2021 & \checkmark     & -          & -  \\
SG \cite{sarlin2020superglue} & 2020 & \checkmark     & -          & - \\
CycleGAN-based$^{*}$ \cite{sindel2022multi}   & 2022 & \checkmark                  & \checkmark         & CycleGAN \cite{zhu2017unpaired}            \\
Content-Adaptive$^{*}$ \cite{wang2021robust} & 2021 & \checkmark & \checkmark & Vessel segmentation \\
\textbf{VDD-Reg (Ours)} & -    & \checkmark                   & \checkmark         & Vessel segmentation \\
\hline 
\end{tabular}%
}
\label{tab_SumMethods}
\begin{minipage}{\columnwidth}
\vspace{0.2cm}
\scriptsize \textit{Deep learning-based} indicates whether the method includes any deep neural network. \textit{Indirect} indicates whether the method applies a transfer method. $^{*}$ indicates that the official implementation is not available, so we implement the method on our own. Some differences might exist, which are discussed in Section~\ref{exp::baseline}. 
\end{minipage}
\end{table}

We compared \textit{VDD-Reg} with five existing methods listed in Table~\ref{tab_SumMethods}. These existing and previously described methods were selected to represent each type of methods we discussed in Sec.~\ref{related:methods}. SURF-PIIFD-RPM \cite{wang2015robust} is one of the well-known traditional methods for multimodal retinal image registration that demonstrated good performance. For the deep-learning based methods, two direct and two indirect methods were chosen for comparison. SuperGlue (SG) \cite{sarlin2020superglue} and LoFTR \cite{sun2021loftr} are two direct methods for feature detection and description which were proposed more recently. For indirect methods, we selected two methods \cite{sindel2022multi, wang2021robust} that utilize different transfer methods, including CycleGAN \cite{zhu2017unpaired} and vessel segmentation.

For fair comparison, we made all methods except for \textit{SURF-PIIFD-RPM} estimate the partial affine transformation matrix and adopt RANSAC with the same hyperparameters. For \textit{SURF-PIIFD-RPM}, as we directly adopted the official code, the affine transformation matrix was applied and RANSAC was not used. Moreover, for methods that used SuperPoint \cite{detone2018superpoint} for keypoint detection and description, including \textit{SG}, \textit{CycleGAN-based}, \textit{Content-Adaptive} and \textit{VDD-Reg}, the official pretrained network was adopted without fine-tuning. More details about these five existing methods are listed as follows:  
\begin{itemize}
    \item \textit{SURF-PIIFD-RPM} \cite{wang2015robust}: This method utilized SURF and PIIFD for more robust feature extraction and RPM for outliers rejection. The official MATLAB code was used.
    \item \textit{LoFTR} \cite{sun2021loftr}: This method exploited Transformer \cite{vaswani2017attention} for processing and matching the dense local features extracted from the backbone. The official pretrained model was adopted with the default setting and was applied directly to the raw images without fine-tuning.
    \item \textit{SuperGlue (SG)} \cite{sarlin2020superglue}: This method used a graph neural network (GNN) for finding correspondences and rejecting non-matchable points between two sets of local features. SuperPoint was used for feature detection and description. The official pretrained networks were adopted, where the SuperPoint detection threshold was set as 0.015 and the SuperGlue match threshold was set as 0.1.
    \item \textit{CycleGAN-based} \cite{sindel2022multi}: This method combined a keypoint detection and description network designed for retinal images (i.e., RetinaCraquelureNet \cite{sindel2022keypoint}) with SuperGlue. The networks were trained using self-supervised learning on synthetic multimodal images generated by CycleGAN \cite{zhu2017unpaired}. As the code was unavailable, we implemented a simplified alternative of this approach by using CycleGAN to transfer images from one modality to another and adopting SuperPoint for feature detection and description.
    \item \textit{Content-Adaptive} \cite{wang2021robust}: This method designed a content-adaptive vessel segmentation network based on the pixel-adaptive convolution (PAC) \cite{su2019pixel} guided by the phase images. The network was trained with style loss and the self-comparison loss. The image registration loss based on the ground truth transformation matrix was also used. We implemented this method based on the code from \cite{zhang2019joint}. Unlike the original paper \cite{wang2021robust}, we ignored the outlier rejection network and did not fine-tuned SuperPoint because these were general techniques applicable to all the other competing methods, which were not within the scope of this paper.
\end{itemize}

\subsection{Training and Testing Details}

All networks in our method were implemented in PyTorch. The network architectures were implemented based on the implementations provided in \cite{zhang2021two} and \cite{detone2018superpoint}, which were adopted with default settings for the DRIU and SuperPoint models, respectively. For the vessel segmentation module, both stages took 1000 epochs for training. The trained networks in stage 1 were used to initialize the networks in stage 2. The Adam optimizer \cite{kingma2014adam} with learning rate 1e-4 was used. A batch size of 1 was used due to the limitation of our GPU memory. OpenCV was used for data pre-processing and the RANSAC algorithm, where the $cv2.estimateAffinePartial2D$ function, an OpenCV function that estimates partial affine transformation, was adopted by setting the maximum reprojection error to 5 pixels and the maximum iteration to 2000.

\subsection{Evaluation Metrics}

\subsubsection{RMSE / MAE}

Based on the predicted registration matrices, the labeled points in all test OCTA images were reprojected to the corresponding test EMA images. Then, the root-mean-square-error (RMSE) \cite{ghassabi2013efficient} between the reprojected points and the labeled points was computed. In addition, the average maximum-absolute-error (MAE) \cite{zhang2021two}, defined as the maximum reprojection error for each image pair, was also reported. 

\subsubsection{Success rate} 

The success rate was defined as the number of image pairs with successful registration over the total number of test image pairs. Since the definition of success varies, we defined a registration as successful when its RMSE $<$ 10 pixels or its MAE $<$ 10 pixels, based on the clinical tolerance.

\subsubsection{Soft Dice / Masked Soft Dice}

\textit{Dice} \cite{li2018multi} is widely used to evaluate the registration quality by calculating the pixel alignment between the warped source vessel masks and the target vessel masks. \textit{Soft Dice} \cite{zhang2021two} has been proposed as an extension of \textit{Dice} for assessing registration quality. For \textit{Soft Dice}, CLAHE \cite{Zuiderveld1994ContrastLA} was first applied to enhance the contrast of two input images and the Frangi vesselness filter \cite{frangi1998multiscale} was then used to generate the vesselness probability masks of the two input images. We calculated \textit{Soft Dice} by
\begin{equation}
Soft\ Dice = \frac{2 \sum_{i=1}^{N} min(F_{i}^{s'}, \ F_{i}^{t})}{\sum_{i=1}^{N} F_{i}^{s'} + \sum_{i=1}^{N} F_{i}^{t}},
\label{eq10}
\end{equation}
where $F^{s'}$ and $F^{t}$ are the vesselness probability masks of the warped source images and the target images, and $N$ denotes the total number of pixels. In our experiments, FA and EMA images were viewed as the source images.

\begin{figure}[htbp]
\centerline{\includegraphics[width=0.9\columnwidth]{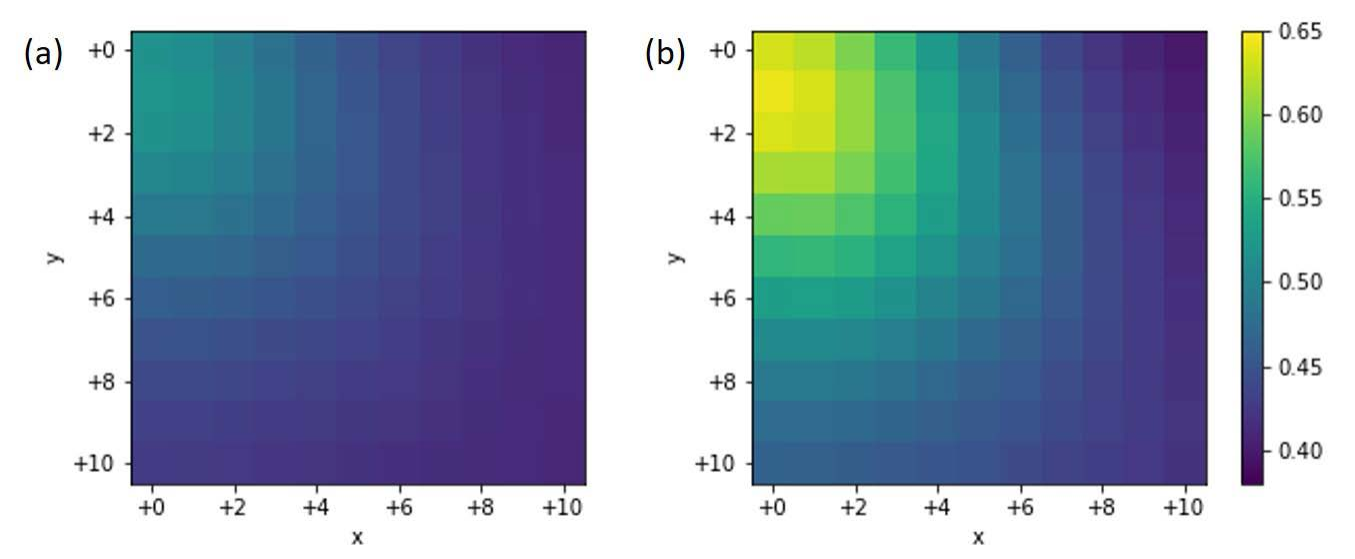}}
\caption{The average (a) \textit{Soft Dice} and (b) \textit{Masked Soft Dice} values over image pairs in \textit{MEMO} by adding different x and y shifts to the ground truth registration. The top-left value in each figure represents the average \textit{Soft Dice} or \textit{Masked Soft Dice} value obtained by ground truth registration. All values are color-coded.}
\label{figSD}
\end{figure}

Additionally, we found that \textit{Soft Dice} could not accurately represent performance when a relatively large VD difference was present between the two modalities, such as in our \textit{MEMO} dataset. Moreover, it was particularly unreliable when the results of each competing method had relatively small differences. As most vessels in OCTA images do not exist in the corresponding EMA images, calculating \textit{Soft Dice} based on every pixel is not ideal. Hence, when evaluating on the \textit{MEMO} dataset, we extended \textit{Soft Dice} to \textit{Masked Soft Dice} which considered only the pixels within the ground truth segmentation masks of EMA images. \textit{Masked Soft Dice} is defined as
\begin{equation}
Masked\ Soft\ Dice = \frac{2 \sum_{i=1}^{N} min(M_{i}^{e'} F_{i}^{e'}, \  M_{i}^{e'} F_{i}^{o})}{\sum_{i=1}^{N} M_{i}^{e'} F_{i}^{e'} + \sum_{i=1}^{N} M_{i}^{e'} F_{i}^{o}},
\label{eq11}
\end{equation}
where $F^{e'}$ and $F^{o}$ are the vesselness probability masks of the warped EMA images and the OCTA images, and $M^{e'}$ represents the warped ground truth segmentation masks of EMA images. In Fig.~\ref{figSD}, we demonstrated that \textit{Masked Soft Dice} has better ability to assess the registration performance on our \textit{MEMO} dataset, as it is less sensitive to the VD difference and noise.

\section{Results}
\label{Result}
\subsection{The CF-FA Dataset}

\begin{table}
\centering
\caption{Results of different methods on the CF-FA test set (Best results are marked in bold)}
\setlength{\tabcolsep}{3pt}
\resizebox{0.8\columnwidth}{!}{
\begin{tabular}{c|c|c|c|c|c}
\hline
Methods & \begin{tabular}[c]{@{}c@{}}Success rate\\ (RMSE \textless{} 10)\end{tabular} & \begin{tabular}[c]{@{}c@{}}Success rate\\ (MAE \textless{} 10)\end{tabular} & RMSE & MAE & Soft Dice \\ \hline
Before Registration & 0.00\% & 0.00\% & 68.35 & 77.03 &  -  \\ 
SURF-PIIFD-RPM \cite{wang2015robust} & 82.76\% & 75.86\% & 15.21 & 20.43 &  0.51  \\
LoFTR \cite{sun2021loftr} & 41.28\% & 27.59\% & 45.13 & 52.29 &  0.51 \\
SG \cite{sarlin2020superglue} & 82.76\% & 79.31\% & 23.68 & 31.40 &  \textbf{0.52} \\
CycleGAN-based (FA$\rightarrow$CF) \cite{sindel2022multi} & \textbf{100.00\%} & \textbf{96.55\%} & 4.86 & 7.52 &  \textbf{0.52}  \\
CycleGAN-based (CF$\rightarrow$FA) \cite{sindel2022multi} & 86.21\% & 75.86\% & 20.09 & 25.77 &  \textbf{0.52}  \\
Content-Adaptive$^{*}$ \cite{wang2021robust} & 89.67\% & 75.86\% &14.03 & 19.27  & 0.51  \\
\textbf{VDD-Reg (Ours)} & \textbf{100.00\%} & \textbf{96.55\%} & \textbf{3.19} & \textbf{5.54} & 0.51 \\
\hline
\end{tabular}}
\label{table_CFFA}
\begin{minipage}{\columnwidth}
\vspace{0.2cm}
\scriptsize $^{*}$The authors of \textit{Content-Adaptive} \cite{wang2021robust} have demonstrated better performance on the CF-FA dataset by fine-tuning SuperPoint and adding an outlier rejection network.  
\end{minipage}
\end{table}

Table~\ref{table_CFFA} illustrates the quantitative results of our method and the existing methods on the CF-FA dataset. The \textit{Masked Soft Dice} metric was not used as the VD difference of the CF-FA dataset is relatively small. From Table~\ref{table_CFFA}, we can observe that our \textit{VDD-Reg} achieved 100\% success rate and the lowest RMSE among all the methods on the CF-FA dataset. Surprisingly, \textit{SURF-PIIFD-RPM}, the only conventional multimodal registration method in Table~\ref{table_CFFA}, achieved decent performance (82.76\%) compared to the other methods. This might indicate that a well-designed conventional method is still competitive if the target dataset is not too difficult. \textit{LoFTR} and \textit{SG} are two \textit{direct} and \textit{deep learning-based} registration methods in Table~\ref{table_CFFA}. \textit{LoFTR}, despite performing well on a general homography estimation dataset \cite{sun2021loftr, balntas2017hpatches}, achieved the worst performance (41.28\%) on the CF-FA dataset according to Table~\ref{table_CFFA}. Fine-tuning \textit{LoFTR} on the CF-FA dataset might be helpful. However, as \textit{LoFTR} was originally trained on the ground-truth labels obtained from a large-scale synthetic indoor scenes datasets \cite{dai2017scannet}, it is unclear how to effectively fine-tune \textit{LoFTR} on a multimodal retinal image registration dataset such as the CF-FA dataset. Compared to \textit{LoFTR}, \textit{SG} demonstrated better generalization to the CF-FA dataset (82.76\%), even though it was trained on the same synthetic dataset \cite{dai2017scannet} as \textit{LoFTR}. This was possibly because SuperPoint (SP), the feature detection and description network used by \textit{SG}, had good generalization capability. Compared to the \textit{direct} methods, \textit{indirect} \textit{deep learning-based} methods in Table~\ref{table_CFFA} generally achieved better registration performance on the CF-FA dataset. \textit{CycleGAN-based (FA$\rightarrow$CF)}, which transformed FA images to CF images with CycleGAN first before registration, also achieved 100\% success rate as our \textit{VDD-Reg}. Its counterpart, \textit{CycleGAN-based (CF$\rightarrow$FA)}, had a slightly lower success rate (86.21\%) mainly because the transformation from FA to CF worked better than the opposite direction using CycleGAN. \textit{Content-Adaptive}, a segmentation-based method similar to our \textit{VDD-Reg}, performed slightly worse (89.67\%) than our method. Due to our two-stage semi-supervised learning framework, \textit{VDD-Seg} could produce vessel segmentation masks more suitable for image registration.

\subsection{The MEMO Dataset}

\begin{figure}[htbp]
\centerline{\includegraphics[width=1.0\linewidth]{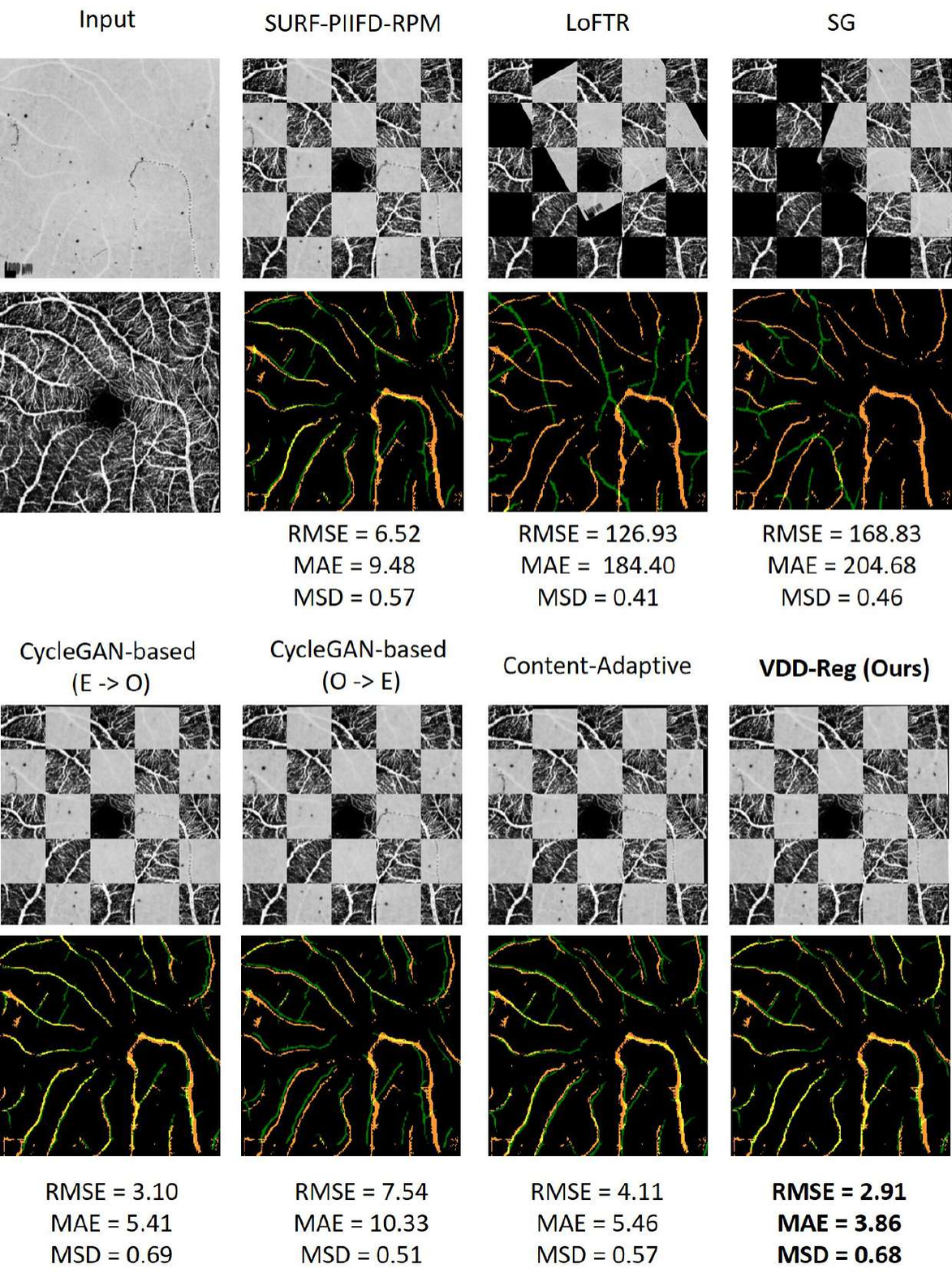}}
\caption{Registration results of our method and the existing methods on a selected image pair from our \textit{MEMO} dataset. The top row shows the grid images where the EMA and OCTA images are interlaced as small grids. The bottom row shows the overlay images of the EMA (green) and OCTA (orange) extracted vessels registered by each method. The vessel segmentation masks were generated by our method. The RMSE, MAE and \textit{Masked Soft Dice} of each method are listed below each overlay image.}
\label{figResults1}
\end{figure}

\begin{table}
\centering
\caption{Results of different methods on the MEMO test set (Best results are marked in bold)}
\setlength{\tabcolsep}{3pt}
\resizebox{1.0\columnwidth}{!}{
\begin{tabular}{c|c|c|c|c|c|c}
\hline
Methods & \begin{tabular}[c]{@{}c@{}}Success rate\\ (RMSE \textless{} 10)\end{tabular} & \begin{tabular}[c]{@{}c@{}}Success rate\\ (MAE \textless{} 10)\end{tabular} & RMSE & MAE & Soft Dice & Masked Soft Dice \\\hline
Before Registration & 6.67\% & 6.67\% & 88.48 & 98.17 & - & - \\
SURF-PIIFD-RPM \cite{wang2015robust} & 6.67\% & 6.67\% & 58.48 & 100.13 & 0.42 & 0.41 \\
LoFTR \cite{sun2021loftr}  & 13.33\% & 13.33\% & 53.55 & 76.37 &  0.43  &  0.49    \\
SG \cite{sarlin2020superglue}   & 0.00\% & 0.00\% & 257.01 & 360.02 & 0.28  &  0.44  \\
CycleGAN-based (OCTA$\rightarrow$EMA) \cite{sindel2022multi} & 33.33\% & 26.67\% & 110.79 & 145.85 &0.42 & 0.51 \\
CycleGAN-based (EMA$\rightarrow$OCTA) \cite{sindel2022multi} & 60.00\% & 40.00\% & 47.96 & 63.74 & 0.47 & 0.56 \\
Content-Adaptive \cite{wang2021robust} & 33.33\% & 20.00\% & 78.69 & 105.03 & 0.41 & 0.48 \\
\textbf{VDD-Reg (Ours)} & \textbf{86.67\%} & \textbf{66.67\%} & \textbf{21.99} & \textbf{28.76} & \textbf{0.50} & \textbf{0.61} \\
\hline
\end{tabular}}
\label{table_MEMO}
\end{table}

Table~\ref{table_MEMO} shows the quantitative registration results of our method and the existing methods on our \textit{MEMO} dataset. We also demonstrate the qualitative registration results of our method and the existing methods on a selected image pair from our \textit{MEMO} dataset in Fig.~\ref{figResults1}. Due to the relatively large VD difference, all the methods performed worse compared to the results on the CF-FA dataset. Still, our \textit{VDD-Reg} outperformed these five existing methods by a large margin (86.67\%). From Table~\ref{table_MEMO}, we observed that \textit{SURF-PIIFD-RPM} performed poorly (6.67\%) on our \textit{MEMO} dataset, which suggested that the hand-crafted features might be insufficiently powerful for this scenario. The two \textit{deep learning-based} \textit{direct} methods, \textit{LoFTR} and \textit{SG}, also produced unsatisfactory results (13.33\% and 0\%) due to the large distribution gap between their training dataset \cite{dai2017scannet} and our \textit{MEMO} dataset. \textit{CycleGAN-based} with pretrained SuperPoint (SP) achieved relatively better performance compared to other competing methods, demonstrating its potential to solving difficult multimodal registration problems. \textit{Content-Adaptive}, which is also a segmentation-based method, performed much worse (33.33\%) than our method on the \textit{MEMO} dataset. We attributed this to our use of annotated vessel segmentation masks from a single modality (EMA in our case). Different from our two-stage semi-supervised learning framework (\textit{LVD-Seg}), \textit{Content-Adaptive} trained the segmentation networks naively with style loss, the self-comparison loss and the image registration loss. To improve the segmentation quality, \textit{Content-Adaptive} additionally guided the segmentation networks with mean phase images of input images using pixel-adaptive convolution (PAC) \cite{su2019pixel}. However, due to the high complexity of OCTA images, the OCTA mean phase images were usually too noisy to correctly guide the segmentation networks. On the other hand, our \textit{LVD-Seg} framework used very few (e.g., three) annotated vessel segmentation masks from one modality to guide the segmentation networks to segment similar vessels in EMA and OCTA image pairs for both stages. In addition, the two-stage design also enhanced training stability when style loss was involved. These are particularly important for multimodal retinal image registration when a large VD difference exists between the two modalities.

\section{Discussion}
\label{Discussion}
\subsection{Ablation Study on the Two-stage Learning Framework}
\label{dis:two-stage}

\begin{table}
\centering
\caption{Results of removing either stage of LVD-Seg when training the segmentation module on the MEMO Dataset (Best results are marked in bold)}
\setlength{\tabcolsep}{3pt}
\resizebox{0.7\columnwidth}{!}{
\begin{tabular}{c|c|c|c|c}
\hline
Method & Success rate & RMSE & Soft Dice & Masked Soft Dice \\ \hline
Stage 1 only & 13.33\% & 108.54 & 0.42 & 0.45 \\
Stage 2 only & 33.33\% & 48.63  & 0.43 & 0.45 \\
Full LVD-Seg & \textbf{86.67\%} & \textbf{21.99} & \textbf{0.50} & \textbf{0.61} \\
\hline
\end{tabular}}
\label{table_difStage}
\end{table}

In this section, we investigated the benefits of each stage in the proposed two-stage semi-supervised learning framework (LVD-Seg) by removing one of the stages from the framework. The results are shown in Table~\ref{table_difStage}. For \textit{Stage 1 only}, we trained the EMA vessel segmentation network following the procedure described in Section~\ref{method:supervised} and used the same EMA vessel segmentation network for segmenting OCTA images. For \textit{Stage 2 only}, we trained the segmentation networks with style loss only. In general, the performance of both variants decreased significantly. \textit{Stage 1 only} performed the worst due to the limited number (three in our case) of annotated vessel segmentation masks used for supervised training, making the segmentation network generalize poorly on the test images and resulting in poor registration performance. Furthermore, \textit{Stage 1 only} directly applied the trained EMA segmentation network on OCTA images, which did not work due to the relatively large VD difference between the two modalities. Although \textit{Stage 2 only} worked better than \textit{Stage 1 only}, it still significantly lagged behind our \textit{LVD-Seg}. This implied that relying purely on style loss could let the training become unstable and resulted in unreliable registration. All these results emphasized the effectiveness of the proposed two-stage semi-supervised learning framework (LVD-Seg).

\subsection{Ablation Study on Number of Required Vessel Masks}
\label{dis:gt}

\begin{table}
\centering
\caption{Results of Using Different Number of Annotated Vessel Segmentation Masks in Stage 1 of LVD-Seg (Best results are marked in bold)}
\label{table1}
\setlength{\tabcolsep}{3pt}
\resizebox{0.7\columnwidth}{!}{
\begin{tabular}{c|c|c|c|c}
\hline
\# of annotated masks & Success rate & RMSE & Soft Dice & Masked Soft Dice \\
\hline
3 (default)  & 86.67\%          & 21.99 & 0.50 & 0.61 \\
5  & 86.67\%           & 11.42 & 0.51 & 0.63 \\
10 & 86.67\%  & \textbf{8.91} & \textbf{0.52} & \textbf{0.65} \\
15 & \textbf{93.33\% } & 11.88 &  \textbf{0.52} & 0.64 \\
\hline
\end{tabular}}
\label{table_DifNum}
\end{table}

The major advantage of our method lies in the requirement for few manually annotated vessel segmentation masks during stage 1 of the \textit{LVD-Seg} framework. In this section, we further investigated the performance of our method on the \textit{MEMO} dataset by using different numbers of labeled EMA vessel segmentation masks for supervised training. Specifically, during stage 1 of \textit{LVD-Seg}, we trained our vessel segmentation module using randomly sampled 3, 5, 10 and 15 annotated EMA vessel segmentation masks. In stage 2, we used all 15 training pairs as our default setting. The results are shown in Table~\ref{table_DifNum}. We found that using more annotated EMA vessel segmentation masks during the supervised training (i.e., stage 1 of \textit{LVD-Seg}) did not affect the performance significantly. For instance, there was only a 6.66\% difference between the highest and the lowest success rates. In other words, the proposed method required very few (e.g., three) annotated vessel segmentation masks to maintain its accuracy, demonstrating its feasibility.

\subsection{Ablation Study on Data Used for Supervised Training}

\begin{table}
\centering
\caption{Results of using different vessel segmentation datasets during stage 1 of LVD-Seg (Best results are marked in bold)}
\setlength{\tabcolsep}{3pt}
\resizebox{0.7\columnwidth}{!}{
\begin{tabular}{c|c|c|c|c}
\hline
Datasets & Success rate & RMSE & Soft Dice & Masked Soft Dice \\
\hline
HRF \cite{budai2013robust}   & 73.33\% & 34.09 & 0.49 & 0.58 \\
DRIVE \cite{staal2004ridge} & 20.00\% & 79.70 & 0.43 & 0.46 \\
MEMO (ours) & \textbf{86.67\%} & \textbf{21.99} & \textbf{0.50} & \textbf{0.61} \\
\hline
\end{tabular}}
\label{table_difPretrain}
\end{table}

As mentioned in the previous section, the primary cost of our method lies in the requirement for few manually annotated vessel segmentation masks. In this section, we further investigated whether existing retinal image segmentation datasets could potentially be used to train our segmentation module during stage 1 of \textit{LVD-Seg}. We selected two datasets, HRF \cite{budai2013robust} and DRIVE \cite{staal2004ridge}, to conduct the experiments. Specifically, HRF and DRIVE are two retinal color fundus (CF) image datasets providing ground truth vessel segmentation masks. We randomly chose three images from each dataset to train our segmentation network during stage 1 of \textit{LVD-Seg}. Other than that, the default settings were adopted. The results are shown in Table~\ref{table_difPretrain}. Compared to the performance of using our \textit{MEMO} dataset, the performance of using the HRF and DRIVE datasets in stage 1 both decreased. Additionally, using the HRF dataset achieved superior performance than using the DRIVE dataset. One possible reason for this was that the HRF dataset had a more similar VD with our \textit{MEMO} dataset. The average VD of \textit{MEMO} (EMA), HRF and DRIVE are 4.71\%, 10.05\% and 11.21\%, respectively. This implies that selecting the ground truth vessel segmentation masks whose VD is closer to the target images (EMA images in our case) might be very important for achieving better results when using the proposed framework.

\subsection{Generalization Ability of the Proposed Method across Different NHPs}
\label{Ablation5}

\begin{table}
\centering
\caption{Results of splitting the dataset by NHPs (Best results are marked in bold)}
\setlength{\tabcolsep}{3pt}
\resizebox{0.85\columnwidth}{!}{
\begin{tabular}{c|c|c|c|c}
\hline
Methods & Success rate & RMSE & Soft Dice & Masked Soft Dice \\
\hline
CycleGAN-based (OCTA$\rightarrow$EMA) \cite{sindel2022multi}   & 18.18\% & 145.90 & 0.31 & 0.40 \\
CycleGAN-based (EMA$\rightarrow$OCTA) \cite{sindel2022multi} & 81.82\% & 27.86 & 0.34 & 0.40 \\
VDD-Reg (Ours) & \textbf{90.90\%} & \textbf{9.38} & \textbf{0.52} & \textbf{0.62} \\
\hline
\end{tabular}}
\label{table_difMonkey}
\end{table}

So far, all experiments conducted on the \textit{MEMO} dataset have utilized a train-test split that included both NHPs in each set. In this section, we further explored whether the proposed method trained on images of one NHP could be generalized to another. To achieve this, we divided the \textit{MEMO} dataset by NHPs. Specifically, eight image pairs captured from one NHP were used as the training set, while the remaining 22 image pairs captured from the other NHP were used as the test set. Other than that, the default settings were adopted for training. The results are shown in Table~\ref{table_difMonkey}. Given that \textit{CycelGAN-based} achieved the second-best performance in Table~\ref{table_MEMO}, we compared our method with it in this study. Interestingly, we found that whether the training and test set included the same NHP did not affect the performance of our method, achieving a success rate of 90.9\%. Additionally, our method outperformed \textit{CycelGAN-based} by 9\% in success rate, indicating its potential to register image pairs from different subjects.

\subsection{Potential of the Proposed Method}
\label{Ablation4}

The proposed \textit{VDD-Reg} requires very little labeling. It could potentially be applied to other vessel imaging modalities, especially for modalities with large differences on vessel structures. This has wider applications for any comparison of SLO images with OCTA. For instance, the registration of FAF to OCTA images may benefit from this approach \cite{narasimha2008registration}. Furthermore, multimodal adaptive optics devices which use both AO-SLO and AO-OCT methods could also benefit from this approach \cite{liu2018trans, le2021novel}.

\section{Conclusion}
\label{DisCon}
In this paper, we present \textit{MEMO}, the first public multimodal EMA and OCTA retinal image dataset. \textit{MEMO} provides registration ground truth, EMA image sequences and OCTA projection images, desirable for various research fields. With \textit{MEMO}, we first uncover a unique challenge of multimodal retinal image registration between modalities with large VD differences. After that, we propose a segmentation-based deep-learning registration framework, \textit{VDD-Reg}, to deal with the large vessel density difference between EMA and OCTA in multimodal retinal image registration. Moreover, to train the segmentation module in our \textit{VDD-Reg}, we design a novel two-stage semi-supervised learning framework, \textit{LVD-Seg}, which combines supervised and unsupervised losses. Both quantitative and qualitative results demonstrate that \textit{VDD-Reg} outperforms the existing methods in both small VD differences (i.e., CF-FA) and large VD differences (i.e., \textit{MEMO}). Additionally, \textit{VDD-Reg} requires as few as three annotated vessel segmentation masks to maintain its performance, which demonstrates its promising potential for registering other modalities. 

\section*{Funding}
The authors would like to acknowledge the funding support from the University of Maryland Strategic Partnership: MPowering the State, a formal collaboration between the University of Maryland College Park and the University of Maryland Baltimore Medical School. This work is also supported by NIH R01EY031731.

\section*{Disclosures} 
The authors declare no conflicts of interest.
\section*{Data availability}
The \textit{MEMO} dataset and CF-FA dataset underlying the results presented in this paper are available at \url{https://chiaoyiwang0424.github.io/MEMO/} and \cite{hajeb2012diabetic}.


\bibliography{refs}






\end{document}